\documentclass[prl,11,twocolumn,floats,aps,epsfig,nofootinbib,amssymb]{revtex4-1}
\usepackage{graphicx}
\usepackage{epsfig}
\usepackage{subfigure}
\usepackage{dcolumn}
\usepackage{bm}
\usepackage{amssymb}
\usepackage{color}
\def\gsim{\lower0.5ex\hbox{$\:\buildrel >\over\sim\:$}}
\def\lsim{\lower0.5ex\hbox{$\:\buildrel <\over\sim\:$}}

\newcommand{\hs}{\hspace*{0.5cm}}

\newcommand{\be}{\begin{equation}}
\newcommand{\ee}{\end{equation}}
\newcommand{\bea}{\begin{eqnarray}}
\newcommand{\eea}{\end{eqnarray}}
\newcommand{\ben}{\begin{enumerate}}
\newcommand{\een}{\end{enumerate}}
\newcommand{\bde}{\begin{widetext}}
\newcommand{\ede}{\end{widetext}}

\newcommand{\al}{\alpha}

\newcommand{\ga}{\gamma}

\newcommand{\fr}{\frac}
\newcommand{\bc}{\begin{center}}
\newcommand{\ec}{\end{center}}

\def\TrTrOneOne{ $\mathrm{ SU(3)_c \otimes SU(3)_L \otimes U(1)_X \otimes U(1)_N}$ }
 
\def\lfv{lepton flavor violation }

\def\vev#1{\left\langle #1\right\rangle}
\newcommand{\sm}{Standard Model }
\begin{document}


\title{ R-parity as a residual gauge symmetry : probing a theory of
  cosmological dark matter}
\author{Alexandre Alves$^1$}
\author{Giorgio Arcadi$^2$}
\author{P. V. Dong$^3$}
\author{Laura Duarte$^4$}
\author{Farinaldo S. Queiroz$^2$}
\author{Jos\'e W. F. Valle$^5$}
\email{queiroz@mpi-hd.mpg.de}

\affiliation{
  $1$ Departamento de F\'isica, Universidade Federal de S\~ao Paulo, Diadema-SP, 09972-270, Brasil\\
  $^2$ Max-Planck-Institut f\"ur Kernphysik, Saupfercheckweg 1, 69117 Heidelberg, Germany\\
  $^3$ Institute of Physics, Vietnam Academy of Science and Technology, 10 Dao Tan, Ba Dinh, Hanoi, Vietnam\\
 $^4$ UNESP - Campus de Guaratinguet\'a -Departamento de F\'isica e Qu\'mica, Guaratinguet\'a - SP - Brazil.\\
  $^5$ AHEP Group, Instituto de F\'isica Corpuscular –
  C.S.I.C./Universitat de Valencia Edificio de Institutos de Paterna,
  C/Catedratico Jos\'e Beltran, 2 E-46980 Paterna (Valencia) - SPAIN }

\begin{abstract}
  We present a non-supersymmetric scenario in which the R-parity
  symmetry $R_P = (-1)^{3(B-L)+2s}$ arises as a result of spontaneous
  gauge symmetry breaking, leading to a viable Dirac fermion WIMP dark
  matter candidate. Direct detection in nuclear recoil experiments
  probes dark matter masses around $2-5$~TeV for
  $M_{Z^{\prime}} \gsim 3-4$~TeV consistent with searches at the LHC,
  while \lfv rates and flavor changing neutral currents in neutral
  meson systems lie within reach of upcoming experiments.
\end{abstract}
\maketitle

\paragraph*{Introduction.---}%

The nature of dark matter is one of the most challenging problems in
science, requiring physics beyond the \sm as well as a new symmetry
capable of making the corresponding particle stable on cosmological
scales.
R-parity is a symmetry imposed by hand in supersymmetry in order to
avoid fast proton decay, leading also to the existence of a stable
Weakly Interacting Massive Particle (WIMP), one of the most compelling
dark matter candidates~\cite{Bertone:2004pz}.
Even if imposed by hand, R-parity may still break through high
dimension operators~\footnote{These can however be forbidden with
  further symmetries \cite{Lee:2010hf,Morisi:2012hu,Dine:2013nga}.  }
or spontaneously~\cite{masiero:1990uj,romao:1992vu}.
While the second case leads to an attractive neutrino mass generation
scheme~\cite{hirsch:2004he}, one loses the WIMP dark matter
scenario~\cite{Restrepo:2011rj}.
Generally, some sort of R-parity symmetry should be invoked in order
to stabilize the dark matter candidate.
For example, an alternative to R-parity in non-supersymmetric schemes
is to impose a discrete lepton number symmetry to stabilize the
WIMP dark matter particle~\cite{Chulia:2016ngi,Chulia:2016giq,Bonilla:2016diq}.\\[-.3cm]

In this work, we discuss a non-supersymmetric model where dark matter
stability results from the R-parity symmetry $R_P = (-1)^{3(B-L)+2s}$,
naturally arising as a consequence of the spontaneous breaking of the
gauge symmetry.
In order to implement this idea we consider an extension of the
standard model based upon an extended \TrTrOneOne electroweak symmetry
broken by Higgs triplets preserving $B-L$. Note that the SU(3)$_L$
symmetry is well-motivated due to its ability to determine the number
of generations to match that of colors by the anomaly cancellation
requirement~\cite{Singer:1980sw,Frampton:1992wt}.
R-parity symmetry $R_P = (-1)^{3(B-L)+2s}$ arises in our model as a
result of spontaneous gauge symmetry breaking, and the stability of
the lightest $R_P$-odd particle leads to a viable Dirac fermion WIMP
dark matter candidate.
We work out the expected rates for direct detection experiments,
flavor changing neutral currents, lepton flavor violation processes
such as $\mu \to e\gamma$, as well as high energy collider signatures.
We also comment on possible connections to cosmological inflation and
leptogenesis.\\[-.2cm]
 
  \paragraph*{The model.--}%
 
  Our non-supersymmetric model is based on the \TrTrOneOne gauge
  group, in which the matter generations are arranged in the
  fundamental representation of $SU(3)_L$ as follows,\\[.1cm]
\begin{tabular}{ccc}
\hline  \hline
Leptons & 1-2nd Generations & 3th Generation\\
$
l_{aL} = \left(\begin{array}{c}
               \nu_{a}\\ e_{a}\\ N_{a}
\end{array}\right)_L $ & $q_{\al L}=\left(\begin{array}{c}
  d_{\al}\\  -u_{\al}\\  D_{\al}
\end{array}\right)_L$ & $q_{3L}=\left(\begin{array}{c} u_{3}\\  d_{3}\\ U \end{array}\right)_L$\\
\\
$\nu_{aR}, e_{aR}, N_{aR}$ & $u_{\al R}, d_{\al R}, D_{\al R}$ & $u_{3R}, d_{3R}, U_{R}$ \\
\hline  \hline
 & Scalars & \\
$\eta =  \left(\begin{array}{c}
\eta^0_1\\
\eta^-_2\\
\eta^0_3\end{array}\right)$ &  $\rho = \left(\begin{array}{c}
\rho^+_1\\
\rho^0_2\\
\rho^+_3\end{array}\right)$ & $\chi = \left(\begin{array}{c}
\chi^0_1\\
\chi^-_2\\ \chi^0_3\end{array}\right),~~~\phi$\\
  \hline
  \hline  \\[.1cm]
\end{tabular}
where we have adopted the generation indices $a=1,2,3$ and $\al=1,2$.

The generators of the Abelian U(1)$_X$ and U(1)$_N$
groups obey the following relations,
\bea Q=T_3-\fr{1}{\sqrt{3}}T_8+X,\hs
B-L=-\fr{2}{\sqrt{3}}T_8+N,\label{ecqbl}
\eea
where $T_i\ (i=1,2,3,...,8)$, $X$ and $N$ are the charges of
$SU(3)_L$, U(1)$_X$ and U(1)$_N$, respectively
\cite{Dong:2013wca,Dong:2014wsa,Dong:2015yra}.
The exotic quarks $U$ and $D$ have electric charge $2/3$ and $-1/3$
respectively.  The quantum numbers associated to the U(1)$_X$ and
U(1)$_N$ symmetries are collected in Table~\ref{tb1}.
\begin{widetext}
\begin{table}[!h]
\begin{center}
\begin{tabular}{|c|cccccccccccccc|}
\hline
Multiplet & $l_{aL}$ & $\nu_{aR}$ & $e_{aR}$ & $N_{aR}$ & $q_{\alpha L}$ & $q_{3L}$ & $u_{aR}$ & $d_{a R}$ & $U_{R}$ & $D_{\alpha R}$ & $\eta$ & $\rho$ & $\chi$ & $\phi$  \\ \hline
$X$ & $-1/3$ & 0 & $-1$ & 0  & 0 & $1/3$ & $2/3$ & $-1/3$ & $2/3$ & $-1/3$ & $-1/3$ & $2/3$ & $-1/3$ & 0 \\ \hline
$N$ & $-2/3$ & $-1$ & $-1$ & 0 & 0 & $2/3$ & $1/3$ & $1/3$ & $4/3$ & $-2/3$ & $1/3$ & $1/3$ & $-2/3$ & 2 \\ \hline
\end{tabular}
\end{center}
\caption{\label{tb1} The $X$ and $N$ charges of the various
  multiplets. Gauge fields have $X=N=0$ and are not listed.}
\end{table}%
 \end{widetext}

Gauge symmetry breaking by these SU(3) Higgs triplets and singlet
addresess the origin of R-parity conservation and dark matter
stability by preserving $B-L$.
Indeed, after the scalar $\phi$ develops a vacuum expectation value
(VEV) at scale $\Lambda$, the continuous U(1)$_N$ symmetry is
spontaneously broken down to the discrete R-parity given as
$ R_P=(-1)^{3(B-L)+2s}=(-1)^{-2\sqrt{3}T_8+3N+2s}$.

We emphasize that this is the only plausible way to embed the $B-L$
symmetry in the model and naturally explain the origin of the
R-parity, since $SU(3)_L$ and $B-L$ symmetries neither commute nor
close algebraically.
We also note that the exotic fermions have the following $B-L$ quantum
numbers, $[B-L](N_a, D_\al,U)=0,-2/3,4/3$, and hence are
$R_P$-odd.
The new Abelian gauge groups give rise to two new neutral gauge bosons
with masses proportional to the B-L and $SU(3)_L$ symmetry breaking
scales, respectively. Unless otherwise stated we will assume that the
B-L symmetry is broken at very high energy scales, implying that only
one new neutral gauge boson, $Z^{\prime}$, will be phenomenologically
relevant. Concerning the exotic quarks they are sufficiently heavy
since their masses are proportional to $w=\vev{ \chi^0_3}$, the VEV of
the $\chi^0_3$ field, taken to be larger than $10$~TeV.

Note that in our model the $N_{aR}$ are truly singlets under the gauge
group, in contrast to the $\nu_{aR}$ which transform under U(1)$_N$,
so they have Dirac masses $h^N_{ab} w$ proportional to
$w=\vev{ \chi^0_3}$.
For all cases, $N_{a}$ can be made the lightest odd particle under
R-parity, and therefore it is a Dirac dark matter candidate (see {\it
  Appendix A} for alternative assumptions). In what follows, we will
investigate the phenomenological consequences of our model. We
start by addressing electroweak limits.\\[-.2cm]
\paragraph*{CKM unitarity.--}%
Quantum loop corrections to the quark mixing matrix resulting from
additional neutral gauge bosons induce deviations from unitarity of
the CKM matrix~\footnote{Here we neglect unitarity violation from
  $\nu_R$
  admixture~\cite{Schechter:1980gr,Escrihuela:2015wra,Miranda:2016ptb}}. These
contributions appear as box-diagrams involving $W$-gauge bosons and
the $Z^{\prime}$ gauge boson leading to hadronic $\beta$-decay, where
the CKM matrix can be extracted from. Such contribution can be
parametrized by
$\Delta_{\mathrm{CKM}} = 1 -\sum_{q=d,s,b} |V_{u,q}|^2$
\cite{Marciano:1987ja}. Applying this to the neutral current we find,

\begin{eqnarray}
\Delta_{\mathrm{CKM}} = -0.0033 \frac{M_{W}^2}{M_{Z'}^2}\ln\left(\frac{M_{W}^2}{M_{Z'}^2}\right) 
\end{eqnarray}which implies into $M_{Z^{\prime}} \gtrsim 200$~GeV.\\[-.2cm]

\begin{table}[htp]
\bc
\begin{tabular}{|c|c|c|}
\hline
$f$ & $g^{f}_V$ & $g^{f}_A$ \\
\hline 
$\nu_a $ & $\fr{ c_{2W}}{2\sqrt{3-4s^2_W}}$ & $\fr{c_{2W}}{2\sqrt{3-4s^2_W}}$ \\
\hline
$e_a$ & $\fr{ 1-4 s^2_W }{2\sqrt{3-4s^2_W}}$ & $\fr{1 }{2\sqrt{3-4s^2_W}}$ \\
\hline 
$N_a$ & $-\fr{ c^2_W}{\sqrt{3-4s^2_W}}$ & $-\fr{ c^2_W}{\sqrt{3-4s^2_W}}$ \\
\hline 
$u_\al$ & $-\fr{3-8s^2_W}{6\sqrt{3-4s^2_W}}$ & $- \fr{1 }{2\sqrt{3-4s^2_W}}$ \\
\hline 
$u_3$  & $\fr{3+2s^2_W}{6\sqrt{3-4s^2_W}}$ & $\fr{ c_{2W}}{2\sqrt{3-4s^2_W}}$ \\
\hline 
$d_\al$ & $-\fr{(3-2s^2_W)}{6\sqrt{3-4s^2_W}}$ & $-\fr{ c_{2W}}{2\sqrt{3-4s^2_W}}$\\
\hline
$d_3$ & $ \fr{\sqrt{3-4s^2_W}}{6}$ & $\fr{1 }{2\sqrt{3-4s^2_W}}$ \\
\hline
\end{tabular}
\caption{\label{tttz2} The couplings of $Z'$ with fermions}
\ec
\end{table}

\paragraph*{Electroweak precision tests.--}%
New physics contributions to the $\rho$-parameter come from the
mixing among the neutral gauge bosons, which is evaluated by,
\be
\Delta\rho\equiv \fr{M^2_W}{c^2_W M^2_{Z_1}}-1\simeq
\fr{(c_{2W}u^2-v^2)^2}{4c^4_W
  (u^2+v^2)w^2}+\fr{t^4_W(u^2+v^2)}{36\Lambda^2},\ee
where $u,\ v,\ w$,\ and $\Lambda$ are the VEVs of
$\eta_1,\ \rho_2,\ \chi_3$, and $\phi$, respectively. Here $Z_1$ is
the lightest of the massive neutral gauge bosons, i.e. the \sm Z boson
in the limit where the scale $w$ is sufficiently high. By enforcing
the experimental limit $\Delta \rho < 0.0006$ \cite{Eidelman:2004wy},
we find the bounds summarized in Table \ref{tabdong}, taking into
account that $u^2+v^2=v_{SM}^2, v_{SM}=246\ \mathrm{GeV}$,
$s^2_W=0.231$, $\al=1/128$, and $g^2=4\pi\al/s^2_W$.
As we  shall see these limits are all surpassed by LHC probes. We
have checked that one-loop new physics corrections to the
$\rho$-parameter dominantly arise from the new non-Hermitian gauge
bosons, but they are much smaller than the tree-level one.
\begin{table}[h]
\begin{tabular}{|l|ccc|}
\hline
$u/v$ & 0 & 1 & $\infty$\\
\hline 
$w$ [TeV] & 6.53 & 1.5 & 3.51\\
$M_{Z^{\prime}}$ [TeV] & 2.58 & 0.593 & 1.4\\
$M_{W^{\prime}}$ [TeV] & 2.12 & 0.494 & 1.14\\
\hline 
\end{tabular}
\caption{\label{tabdong} Bounds on the $w$ symmetry breaking scale
  from electroweak measurements on the $\rho$-parameter. This bound
  can be translated into limits on the gauge boson masses
  $M^2_{Z^{\prime}}\simeq \frac{g^2c^2_W w^2}{3-4 s^2_W}$ and
  $M_{W^{\prime}}^2=\frac{g^2( v_{SM}^2 +w^2)}{4}$. }
\end{table}
       

A more robust bound, insensitive to the VEV hierarchy, stems from the
amazing precision achieved by LEP. This still provides a good test for
new neutral gauge bosons that couple to leptons via the
$e^+e^- \to Z^{\prime} \to f \bar{f}$ production channel with
$Z^{\prime}$ being off-shell. The bound can be obtained using the
parametrization \cite{LEP:2003aa},
\begin{eqnarray}
 \mathcal{L} =\fr{g^2}{c^2_W M^2_{Z^{\prime}}}[\bar{e}\ga^\mu(a^e_LP_L+a^e_R P_R)e][\bar{f}\ga_\mu(a^f_L P_L+a^f_R P_R f]\nonumber\\
\end{eqnarray}where $a^f_L=(g^f_V+g^f_A)/2$ and $a^f_R=(g^f_V-g^f_A)/2$.
%
%

Using the LEP-II results for final state dileptons we get
\cite{Carena:2004xs},
\begin{eqnarray}
 \frac{g^2}{\cos^{2}(\theta_{W})}\left[\frac{\cos(2\theta_{W})}{2\sqrt{3-4\sin^2(\theta_{W})}}\right]\frac{1}{M_{Z'}^2}<\frac{1}{{\rm (6\,T\,eV)^2}}
\end{eqnarray}
which translates into $M_{Z^{\prime}}> 1.93$~TeV. \\[-.2cm]
\paragraph*{Muon Magnetic Moment.--}%
Any fundamental charged particle has a magnetic dipole moment ($g$)
which is parametrized in terms of $a = (g-2)/2$. In the case of the
electron the SM prediction agrees quite well with the experimental
observation, constituting a capital example of the success of quantum
field theory. On the other hand, for the muon, there is a long
standing discrepancy between theory and measurement of about
$3.6\sigma$ \cite{Blum:2013xva}.
This translates into $\Delta a_{\mu} =(287 \pm 80)\times 10^{-11}$
(see \cite{Lindner:2016bgg} for an extensive and recent review).
The ongoing $g-2$ experiment at FERMILAB will shed light into this
problem and, should the central value remain intact, a $5\sigma$
evidence for new physics would result, with
$\Delta a_{\mu} =(287 \pm 34)\times 10^{-11}$.
The model presented here cannot account for $g-2$, since the required
gauge boson masses would be too small to fulfill current experimental
limits from high energy colliders (see below). Hence one can only
derive limits, by requiring their contribution to lie within the error
bars. Using the current (projected) sensitivities we find,
$M_{Z^{\prime}} > 180\, \mathrm{GeV}\ (273\, \mathrm{GeV}),
M_{W^{\prime}} > 100\, \mathrm{GeV}\ (145\, \mathrm{GeV})$.\\[-.2cm]
\paragraph*{Flavor changing neutral current.--}%
Mesons are unstable systems, but if their lifetime is sufficiently
long we can observe them at colliders. The $K^0$ meson, a bound state
of $d\bar{s}$, is necessarily different from its antiparticle due to
strangeness. As a result of CP violation in weak interactions, these
mesons decay differently, and their mass difference has been used as a
sensitive probe for flavor changing interactions.
Similar discussion holds for the $B^0_s -\bar{B}^0_s$ meson
system. Defining $V_{\mathrm{CKM}}= U_L^{\dagger}V_L$, with
$U_L (V_L)$ being the matrix relating the flavor states to the
mass-eigenstates of positive (negative) isospin, one can find that the
contribution to the mass difference for meson systems is,
\be
K^{0}-\bar{K^{0}}: \frac{g^2}{(3-t^2_W)M_{Z'}^2
}[(V^{*}_{L})_{31}(V_{L})_{32}]^2 < \frac{1}{(10^4\,\mathrm{TeV})^2},
\ee
\be B_{s}^{0}-\bar{B_{s}^{0}}: \frac{g^2}{(3-t^2_W)
  M_{Z'}^2}[(V^{*}_{L})_{32}(V_{L})_{33}]^2 <
\frac{1}{(100\,\mathrm{TeV})^2}.  \ee

The bound derived on the mass of the $Z^{\prime}$ gauge boson is
rather sensitive to the parametrization used for the $V$ matrix that
diagonalizes the CKM matrix.
In \cite{Queiroz:2016gif} two possible parametrizations were
considered, that yield either optimistic or conservative limits, while
keeping the CKM matrix in agreement with data. In the optimistic one,
one finds $V_{31}=0.43$, $V_{32}=0.089$, $V_{33}=0.995$, with the the
$K-\bar{K^0}$ system producing the strongest limit,
$M_{Z^{\prime}} > 14$ TeV.
Taking a conservative approach, one finds $V_{31} = 0.00037$,
$V_{32} = 0.052$, $V_{33} = 0.998$, with the $B^0_s-\bar{B}^0_s$
system offering a better probe, implying the lower bound
$M_{Z^{\prime}} > 1.95$~TeV. Thus it is clear that meson systems can
be powerful tests for new physics effects, although suffer from
important uncertainties. In this work we will adopt the conservative
bound, but bear in mind that more stringent limit may be applicable.\\[-.2cm]
\paragraph*{Dilepton Resonance Searches at the LHC.--}%
%
Dilepton resonance searches are the gold channel for heavy neutral
gauge bosons with un-suppressed couplings to leptons
\cite{ATLAS2}. Since signal events are peaked at the $Z^{\prime}$
mass, the use of cuts on the dilepton invariant mass is a powerful
discriminating tool.  The background comes mainly from Drell-Yann
processes and is well understood
\cite{Jezo:2015rha,Klasen:2016qux}. Using 13 TeV center-of-energy and
$3.2{\rm fb^{-1}}$ of integrated luminosity the ATLAS collaboration
has placed restrictive limits on the mass of gauge bosons arising in
some new physics models \cite{Aad:2014cka}. Using the just released
LHC data with integrated luminosity of
$\mathcal{L}=13.3{\rm fb^{-1}}$, and applying the cuts,
\begin{itemize}
\item  $E_T(e_1) > 30 \,{\rm GeV}$, $E_T(e_2) > {\rm 30 \,GeV}$, $|\eta_e| < 2.5$,
\item  $p_T(\mu_1) > 30 \,{\rm GeV}$, $p_T(\mu_2) > 30 \,{\rm GeV}$, $|\eta_{\mu}| < 2.5$,
\item $500 \,{\rm GeV} < M_{ll} < 6000 \,{\rm GeV}$,
\end{itemize}
with $M_{ll}$ denoting the dilepton invariant mass, one can find a
bound on the $Z^{\prime}$ mass \footnote{See \cite{Queiroz:2016gif}
  and \cite{Barreto:2010ir,Coutinho:2013lta,Salazar:2015gxa} for
  previous studies.}.
We have generated events with MadGraph5
\cite{Alwall:2014hca,Alwall:2011uj}, adopting the CTEQ6L parton
distribution function \cite{Lai:2009ne} and efficiencies/acceptances
as described in \cite{Aad:2014cka}. The resulting limit was found to
be $M_{Z^{\prime}} > 3.8$~TeV.
Keeping a similar detector response we expect that upcoming LHC runs
with $\mathcal{L}=100 (1000)\, {\rm fb^{-1}}$ will probe
$M_{Z^{\prime}}=4.9\, (6.1)$~TeV, respectively.\\[-.2cm]
\paragraph*{Charged Lepton + MET at the LHC.---}%
%
The presence of a charged gauge boson ($W^{\prime}$) is a feature
shared among all models based on the $SU(3)_L$ gauge group.
In order to constrain the mass of this charged gauge boson one looks
for high transverse mass signal events~\cite{Aaboud:2016zkn,ATLAS3}.
Here one can use the lepton plus missing energy data, via the
$pp \to W^{\prime} \to l\nu$ production channel at the LHC with
$\mathcal{L}=13.3fb^{-1}$ and 13 TeV center of mass energy.
No significant excess above \sm predictions was seen, leading to
$M_{W^{\prime}} > 4.74$~TeV \cite{ATLAS3}.  In this model, the charged
current contains,
\begin{equation}
\mathcal{L} \supset -\frac{g}{\sqrt{2}} \left( \bar{e}_{aL}\gamma^{\mu} N_{aL} \right) W^{\prime}_\mu.
\label{Eq:chargedcurrent}
\end{equation}

\begin{figure}[h]
 \centering
 \includegraphics[width=\columnwidth]{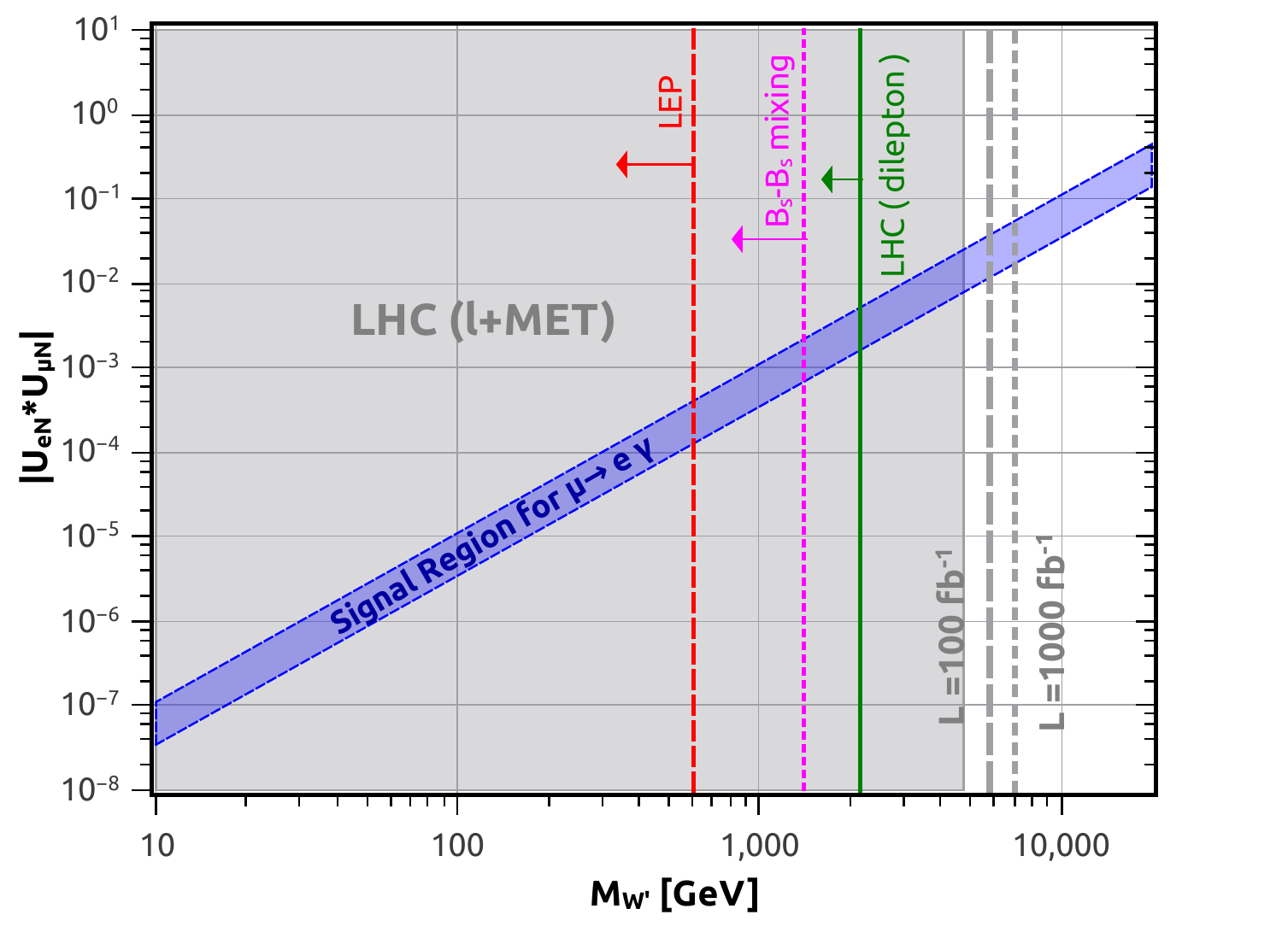}
 \caption{Region of parameters yielding
   $4.2 \times 10^{-13} < {\rm Br(\mu \to e\gamma)} < 4 \times
   10^{-14}$ in blue, overlaid with bounds from LEP (dashed red),
   $B_s^0 -\bar{B}^0_s$ mixing (dashed pink), dilepton data from LHC
   (solid green), and l+MET data from LHC in gray. The upper blue line
   in the region represents the current limit
   ${\rm Br(\mu \to e\gamma)} < 4.2 \times 10^{-13}$.}
 \label{fig2}
\end{figure}

Since this charged gauge boson will be assumed to be much heavier than
the lightest $N$ (i.e. odd fermion in our case), we expect that the
signal events will have approximately the same cut efficiencies
observed in the ATLAS study.
Given that the interactions of the Lagrangian in
Eq. (\ref{Eq:chargedcurrent}) is similar to the one considered in
$W^{\prime}$ searches, the bound above is expected to be applicable to
our model. This limit is represented by the gray region in
Fig. \ref{fig2}. We also looked at the prospects for future runs from
the LHC at $13$~TeV, with $\mathcal{L}=100 (1000)\, {\rm fb^{-1}}$ which turn out
to be sensitive to $M_{W^{\prime}}=5.8\, (7)$~TeV.
\paragraph*{Lepton flavor violation.--}%
%
In the \sm lepton flavor is conserved and neutrinos are
massless. However, neutrinos experience flavor oscillations
\cite{Forero:2014bxa,Kajita:2016cak,McDonald:2016ixn} which is a
direct confirmation that leptonic flavor is violated.
An observation of charged \lfv would have enormous impact on
our understanding of the lepton sector and could have important
implications for new physics.
Indeed, the existence of \lfv in neutrino propagation suggests that it
should also exist in the charged lepton sector, leading to decays such
as $\mu \to e\gamma$.
Unfortunately the connection is highly
model-dependent~\cite{bernabeu:1987gr}.
In our model, the presence of right-handed neutrinos (i.e. odd
fermions), with the lightest one constituting the dark matter, can
mediate a fast decay $\mu \to e\gamma$ via $W^{\prime}$ exchange, with
a branching ratio found to be \cite{Lindner:2016bgg},
\begin{eqnarray}
{\rm Br(\mu \to  e\gamma)} = 6.43\times 10^{-6}\left(\frac{1\ \mathrm{TeV}}{M_{W'}}\right)^4\sum\limits_{f}(g^{fe*}g^{f\mu})^2,\nonumber\\
\end{eqnarray}with $g^{fe}=g\,U^{Ne\ast}/(2\sqrt{2})$ and $g^{f\mu}=g\,U^{N\mu\ast}/(2\sqrt{2})$.\\[-.2cm]

\begin{figure}[!h]
 \centering
 \includegraphics[width=0.8\columnwidth]{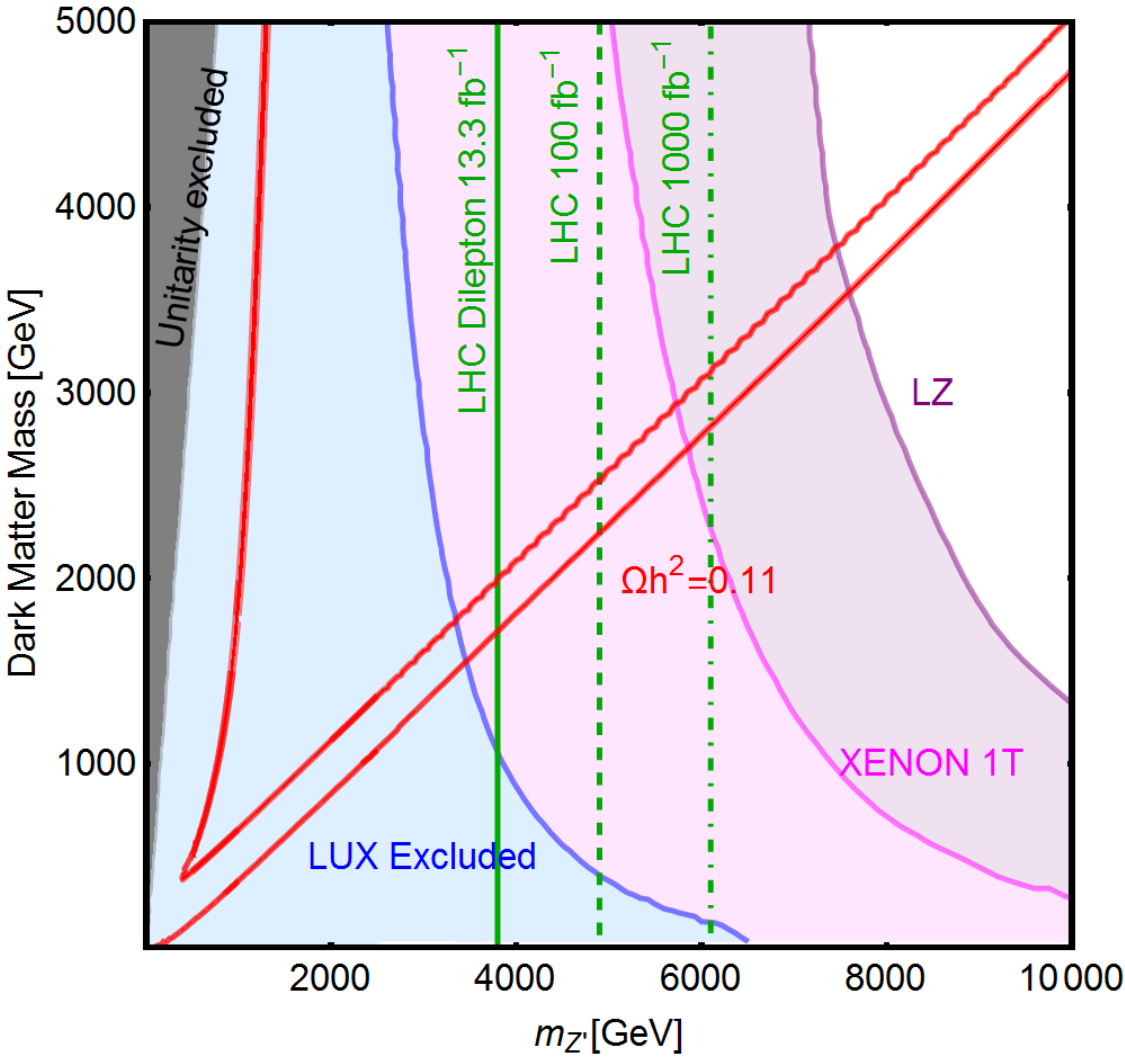}
 \caption{Region of parameters that yields the right relic density
   curve in red. The existing limits from the non-observation of dark
   matter matter-nucleon scattering by the LUX collaboration are
   indicated in light blue \cite{Aprile:2015uzo}. The prospects for
   the XENON1T experiment with 2-years exposure~\cite{Aprile:2015uzo},
   as well as the projected sensitivy of the LZ dark matter experiment
   are also indicated~\cite{McKinsey:2016xhn}.  Current limits as well
   as projected sensitivities from LHC searches of dilepton resonances
   for luminosities of $100\,{\mbox{fb}}^{-1}$ and
   $1000\,{\mbox{fb}}^{-1}$ are also shown.}
 \label{fig3}
\end{figure}

Current (projected) sensitivity as reported by the MEG collaboration
\cite{TheMEG:2016wtm} implies that
${\rm Br(\mu \to e\gamma)} < 4.2 \times 10^{-13}\ (4 \times
10^{-14})$.
Thus one can translate this bound into a limit on the product
$U^{Ne\ast}U^{N\mu}$ as function of the $W^{\prime}$ mass as shown in
Fig.\ref{fig2}. There we have overlaid the aforementioned constraints
altogether as indicated in the caption. There we have converted the limits on the $Z^{\prime}$ mass into bounds on the $W^{\prime}$ knowing that their mass are determined by a common energy scale w. We conclude that depending on
the value for the product $U^{Ne\ast}U^{N\mu}$ the $\mu \to e\gamma$
search may outperform collider probes. We now investigate the
feasibility of this
model concerning dark matter searches.\\[-.2cm]
\paragraph*{WIMP Dirac dark matter.---}%
%
In our model, one can have either a Dirac or Majorana fermionic dark
matter~\cite{Mizukoshi:2010ky,Alvares:2012qv,Profumo:2013sca,Kelso:2013nwa,Cogollo:2014jia,Kelso:2014qka,Queiroz:2016sxf}, though in this work we focus on the
Dirac possibility, since the Majorana case is already excluded by
combining the existing constraints ({\it see Appendix A}).
As we discussed earlier the dark matter mass can be regarded as a free
parameter. The current dark matter relic density and scattering rate
at underground detectors are dictated, respectively, by the s-channel
and t-channel $Z^{\prime}$-induced interactions.
The $W^{\prime}$ boson also mediates t-channel interactions, which are
nevertheless subdominant, and thus neglected in our computations.
In Fig. \ref{fig3} we exhibit the result of the dark matter
phenomenology, encompassing the relic density using PLANCK data
\cite{Ade:2015xua}, dark matter-nucleon scattering limits from
searches by the LUX collaboration \cite{Akerib:2016vxi}, expected
XENON1T \cite{Aprile:2015uzo} and LZ \cite{McKinsey:2016xhn}
sensitivities, as well as current and future prospects at the LHC for
searches for neutral gauge bosons.
It is clear that current limits from the LHC and dark matter direct
detection are rather complementary; in particular LHC can test the
WIMP paradigm for higher values of the dark matter mass. It is
nevertheless exciting to observe that next generation direct detection
experiments, i.e. XENON1T and LZ, are expected to probe the model for
$Z^{\prime}$ mases up to $10$~TeV outperforming the LHC. In conclusion
we have shown that this UV complete dark matter model addresses the
origin of R-parity from first principles and offers a viable dark
matter candidate for $M_{Z^{\prime}} > 4$~TeV and dark matter masses
of $2-5$~TeV.

\paragraph*{Conclusion.---}%

Is summary, we have presented a non-supersymmetric \TrTrOneOne model
in which a conserved R-parity symmetry $R_P = (-1)^{3(B-L)+2s}$ arises
as a residual unbroken discrete gauge symmetry.
The fact that the $B-L$ symmetry is preserved at high scales plays a
key role in accounting for the origin of R-parity conservation.
The lightest $R_P$-odd particle constitutes a viable WIMP dark matter
candidate, whose stability follows naturally from the breaking of the
gauge symmetry.
We have shown that the scheme offers good prospects for dark matter
detection in nuclear recoil experiments, as well as flavor changing
neutral currents in the neutral meson systems $K^{0}-\bar{K}^{0}$ and
$B^0-\bar{B^0}$, searches for lepton flavor violating processes such
as $\mu \to e\gamma$, as well as dilepton event searches at the LHC.

\paragraph*{Acknowledgments.---}%
A.A acknowledges support from Brazilian agencies CNPq (process 307098/2014-1), and FAPESP (process 2013/22079-8);JWFV from Spanish grants FPA2014-58183-P, Multidark
CSD2009-00064, SEV-2014-0398 (MINECO) and PROMETEOII/2014/084 (GVA); P.V.D from
Vietnam National Foundation for Science and Technology
Development (NAFOSTED) under grant number 103.01-2016.77; and L.D from CAPES.

\appendix

\section{Appendix A: Inviability of Majorana dark matter}%

In the model discussed thus far the neutral fermions have a
Lagrangian term $h^N \bar{l}_L \chi N_R $
%
%
where $h^N$ is the relevant Yukawa coupling.
After $\chi$ develops a nonzero VEV $\vev{\chi} = (0,0,w)/\sqrt{2}$
one obtains three heavy Dirac fermions $N$, with masses at the large
symmetry breaking scale $\vev{\chi}$.
Notice however, that one can also add a bare mass term
$ N_R N_R$ proportional to a mass parameter $\mu$.
For $\mu \to 0$ we have Dirac fermions, while for $\mu \ll w$ the
global symmetry enforcing Diracness is only approximate, and the
$N_R$ become quasi-Dirac fermions~\cite{valle:1982yw}.
On the other hand, for arbitrary $\mu \sim w$ they are generic
Majorana fermions.

\begin{figure}[!h]
\centering
\includegraphics[width=0.8\columnwidth]{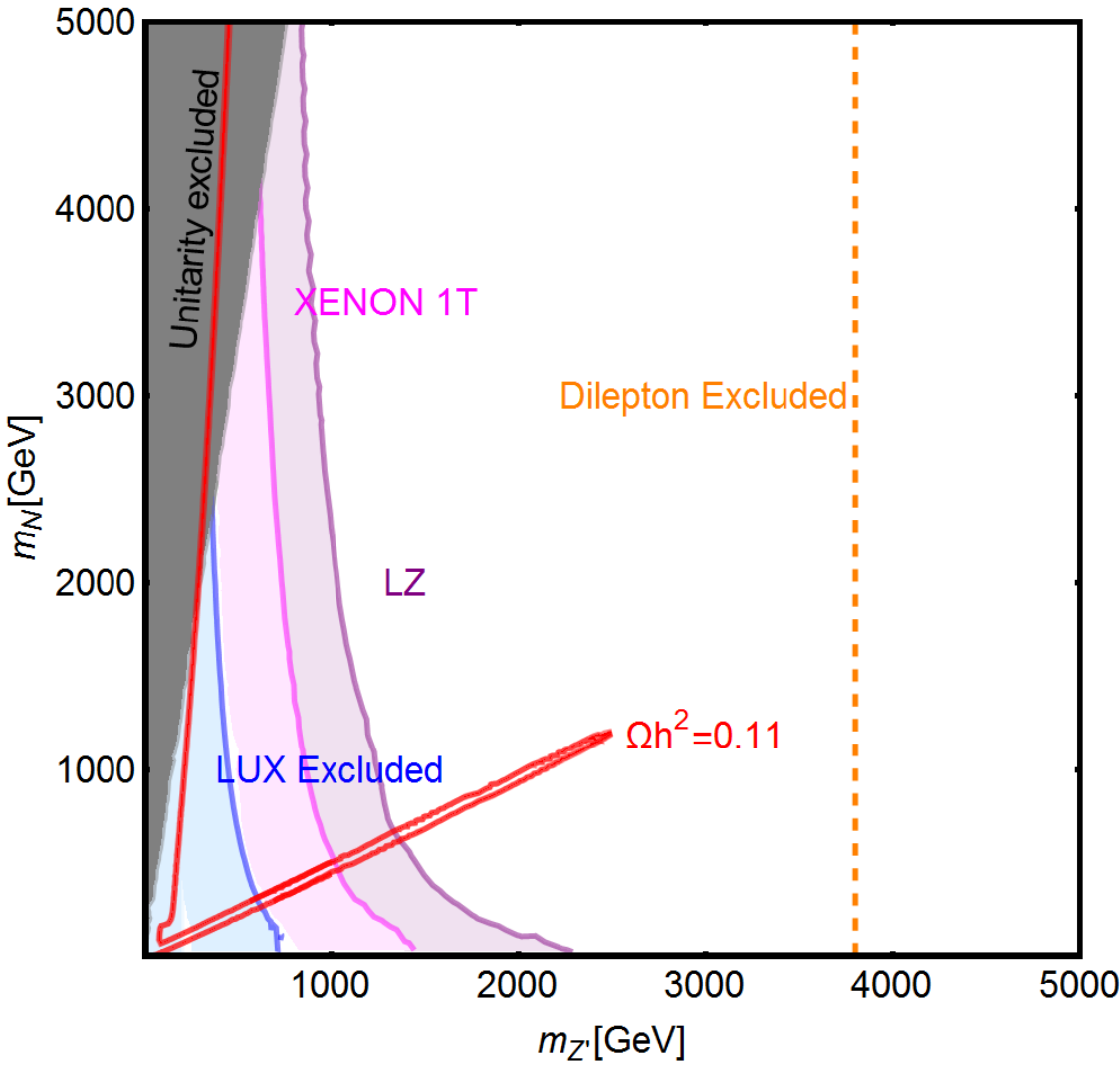}
\caption{Inviability of Majorana dark matter: the plot shows how
  existing dilepton event search limits preclude a viable Majorana
  dark matter candidate.}
 \label{figappendix1}
\end{figure}

Such a model would be perfectly consistent except that the dark matter
interpretation would no longer be viable.
Indeed, if the lightest of the $N_a$ is a Majorana fermion its
vectorial coupling with the $Z^{\prime}$ gauge boson vanishes,
affecting its relic density calculation as well its direct detection
rates.  In Fig.\ref{figappendix1} we show the final result of having
$N_a$ as Majorana fermions after implementing collider and
spin-dependent direct detection limits.  One sees that a Majorana dark
matter fermion is already excluded in view of the current limits.

This highlights the testability of the model, since the couplings are
all fixed by gauge symmetry.  Therefore, the bare mass term must be
suppressed, making the $N_a$ (mainly) Dirac fermions and restoring the
results discussed the main text.

\section{Appendix B: Neutrino seesaw mechanism,
  leptogenesis and cosmological inflation}

Here we note that the neutrinos have Yukawa Lagrangian terms given by
\be
\mathcal{L}\supset h^\nu_{ab} \bar{l}_{aL} \eta \nu_{bR} + \frac 1 2
f^\nu_{ab} \bar{\nu}^c_{aR} \phi \nu_{bR} +H.c.
\ee
After the scalars develop nonzero vacuum expectation values,
$\vev{\eta} = (u, 0, 0)/\sqrt{2}$ and $\vev{\phi} = \Lambda/\sqrt{2}$,
this leads to $m_\nu \simeq - m_D m^{-1}_R m^T_D\sim u^2/\Lambda$,
where $m_D=-h^\nu u/\sqrt{2}$.  Since $\Lambda \gg u$ the light
neutrino masses are naturally small thanks to the canonical type-I
seesaw mechanism.
On the other hand the heavy right-handed neutrinos $\nu_R$ have large
Majorana masses $m_R=-f^\nu\Lambda/\sqrt{2}$, at the U(1)$_N$ breaking
scale~\footnote{For the systematic seesaw expansion formalism
  see~\cite{Schechter:1981cv}}.

Here we note that the U(1)$_N$ breaking that defines R-parity can not
only lead to neutrino masses, but also potentially lead to
leptogenesis, and cosmological inflation, in certain correlations to
dark matter.

The leptogenesis mechanism is governed by CP-asymmetric $\nu_R$ decays
which, besides the usual mode $\nu_R\to e\eta_2$ include a new mode
$\nu_R\to N\eta_3$ to $R_P$-odd states. This channel transforming into
the dark sector is enhanced with respect to the former.
We expect there is a link between the fermionic dark matter and the
matter-antimatter asymmetry. Note also that the effective potential of
$\phi$ which has $\nu_R$, Higgs triplets, and U(1)$_N$ gauge field
contributions, may easily satisfy slow-roll conditions required for
cosmic inflation. Inflation would end seemingly due to an instability
triggered by $\phi$ when it reaches a critical value defined by the
largest Higgs triplet mass. The inflaton eventually decays into odd
scalars, $\phi\to \eta_3\eta_3$, while the channel into fermionic dark
matters is loop-induced. This would ensure that fermionic dark matter
particles are thermally produced.

\bibliography{combined}

\end{document}